\input harvmac
\noblackbox

\font\tenbifull=cmmib10 
\font\tenbimed=cmmib10 scaled 800
\font\tenbismall=cmmib10 scaled 666
\textfont9=\tenbifull \scriptfont9=\tenbimed
\scriptscriptfont9=\tenbismall

\def\valpha{{\fam=9{\mathchar"710B } }}
\def\vbeta{{\fam=9{\mathchar"710C } }}
\def\vn{{\bf n}}
\def\vh{ {\bf h}}
\def\vq{ {\bf q}}
\def\vH{ {\bf H}}
\def\vg{ {\bf g}}
\def\p{\partial}
\def\inbar{\,\vrule height1.5ex width.4pt depth0pt}
\def\IB{\relax{\rm I\kern-.18em B}}
\def\IC{\relax\hbox{$\inbar\kern-.3em{\rm C}$}}
\def\ID{\relax{\rm I\kern-.18em D}}
\def\IE{\relax{\rm I\kern-.18em E}}
\def\IF{\relax{\rm I\kern-.18em F}}
\def\IG{\relax\hbox{$\inbar\kern-.3em{\rm G}$}}
\def\IH{\relax{\rm I\kern-.18em H}}
\def\II{\relax{\rm I\kern-.18em I}}
\def\IK{\relax{\rm I\kern-.18em K}}
\def\IL{\relax{\rm I\kern-.18em L}}
\def\IM{\relax{\rm I\kern-.18em M}}
\def\IN{\relax{\rm I\kern-.18em N}}
\def\IO{\relax\hbox{$\inbar\kern-.3em{\rm O}$}}
\def\IP{\relax{\rm I\kern-.18em P}}
\def\IQ{\relax\hbox{$\inbar\kern-.3em{\rm Q}$}}
\def\IR{\relax{\rm I\kern-.18em R}}
\font\cmss=cmss10 \font\cmsss=cmss10 at 7pt
\def\IZ{\relax\ifmmode\mathchoice
{\hbox{\cmss Z\kern-.4em Z}}{\hbox{\cmss Z\kern-.4em Z}}
{\lower.9pt\hbox{\cmsss Z\kern-.4em Z}}
{\lower1.2pt\hbox{\cmsss Z\kern-.4em Z}}\else{\cmss Z\kern-.4em Z}\fi}

%
%

\def\NP{{\it Nucl. Phys.\ }}

\def\PL{{\it Phys. Lett.\ }}
\def\PR{{\it Phys. Rev.\ }}

\def\CMP{{\it Comm. Math. Phys.\ }}

\def\PRep{{\it Phys. Rep.\ }}
%
%
\font\tiau=cmcsc10
%
%
\lref\wit{E. Witten,
\PL {\bf 86B} (1979)
283.}
\lref\gno{P. Goddard, J. Nuyts and D. Olive,
\NP {\bf B125} (1977) 1.}
\lref\engwind{F. Englert and P. Windey, Phys. Rev. {\bf D14} (1976) 2728.}
\lref\lwy{K. Lee, E. Weinberg and P. Yi, ``Electromagnetic Duality and
SU(3) Monopoles", Columbia Preprint CU-TP-734,  hep-th/9601097.}
\lref\con{S. A. Connell, ``The Dynamics of the $SU(3)$ Charge $(1,1)$
Magnetic Monopole", University of South Australia Preprint.}
\lref\godol{P. Goddard and D. Olive, \NP {\bf B191} (1981) 511.}
\lref\corol{E. Corrigan, D. I. Olive, D. B. Fairlie and J. Nuyts, ``Magnetic
Monopoles in $SU(3)$ Gauge Theories,'' \NP {\bf B106} (1976) 475.}
\lref\taubes{C. H. Taubes, \CMP {\bf 81} (1981) 299.}
\lref\horraw{P. A. Horvathy and J.H. Rawnsley,
\CMP {\bf 96} (1984) 497; \CMP {\bf 99} (1985) 517.}
\lref\brown{L. S. Brown, R.D. Carlitz and C. Lee, \PR {\bf D16} (1977) 417.}
\lref\brill{D. Brill, Phys. Rev. {\bf B133} (1964) 845.}
\lref\pope{C. N. Pope, Nucl. Phys. {\bf B141} (1978) 432.}
\lref\cvetic{M. Cvetic and D. Youm, ``Dyonic BPS Saturated Black Holes of
Heterotic String on a Six Torus", hep-th/9507090.}
\lref\mottola{E. Mottola, \PL {\bf 79B} (1978) 242; \PR {\bf D19} (1979) 3170.}
\lref\weinberg{E. J. Weinberg,
\NP {\bf B167} (1980) 500.}
\lref\sen{A. Sen,
\PL {\bf B329} (1994) 217.}
\lref\mo{C. Montonen and D. Olive, Phys. Lett. {\bf 72B} (1977) 117.}
\lref\osborn{H. Osborn, Phys. Lett. {\bf 83B} (1979) 321.}
\lref\girard{L. Girardello, A. Giveon, M. Porrati and A. Zaffaroni,
hep-th/9406128, \PL {\bf B334} (1994) 331; hep-th/9502057,
\NP {\bf B448} (1995) 127.}
\lref\listnew{C. Vafa and E. Witten, hep-th/9408074, \NP {\bf B431} (1994) 3;
J.A. Harvey, G. Moore and A. Strominger, hep-th/9501022;
M. Bershadsky, A. Johansen, V. Sadov and C. Vafa,  hep-th/9501096,
\NP {\bf B448} (1995) 166;
N. Dorey, C. Fraser, T. Hollowood and M. A. C. Kneipp,
``Non-Abelian Duality in N=4 Supersymmetric Gauge Theories", hep-th/9512116.}
\lref\porrati{M. Porrati, ``On the Existence of States Saturating the
Bogomol'nyi Bound in N=4 Supersymmetry", hep-th/9505187.}
\lref\sethietal{S. Sethi, M. Stern and E. Zaslow,
``Monopole and Dyon Bound States in $N=2$ Supersymmetric Yang-Mills Theories",
hep-th/9508117.}
\lref\gauntharv{J. P. Gauntlett and J. Harvey ``$S$-Duality and the Dyon
Spectrum in $N=2$ Super Yang-Mills Theory", CALT-68-2017, hep-th/9508156.}
\lref\gauntlett{J. P. Gauntlett, \NP {\bf B411} (1994) 443.}
\lref\segal{G. Segal, talk given at the Newton Institute, Cambridge,
Summer 1994,
unpublished.}
\lref\blum{J. D. Blum, \PL {\bf B333} (1994) 92.}
\lref\bais{F. A. Bais, \PR {\bf D18} (1978) 1206.}
\lref\atiyah{M. F. Atiyah and N. J. Hitchin, ``The Geometry and Dynamic of
Magnetic Monopoles,'' Princeton Univ. Press, Princeton, NJ, 1988.}
\lref\slansky{R. Slansky, 
\PRep {\bf 79} (1981) 1.}
%
%
\baselineskip 12pt
\Title{\vbox{\baselineskip12pt
\hbox{CALT-68-2035}
\hbox{UCSBTH-95-36}
\hbox{\tt hep-th/9601085} }}
{\vbox{\hbox{\centerline{\bf DYONS AND S-DUALITY IN N=4 SUPERSYMMETRIC
GAUGE THEORY
}}}}
\centerline{\tiau Jerome P. Gauntlett}
\centerline{\it California Institute of Technology, Mail Code 452-48}
\centerline{\it Pasadena, CA 91125}
\centerline{\it jerome@theory.caltech.edu}
\centerline{\tiau and}
\centerline{\tiau David A. Lowe}
\centerline{\it Department of Physics}
\centerline{\it University of California}
\centerline{\it Santa Barbara, CA 93106-9530}
\centerline{\it lowe@tpau.physics.ucsb.edu}
\vskip .5cm
\centerline{\bf Abstract}
We analyze the spectrum of dyons
in $N=4$ supersymmetric Yang-Mills theory with gauge group $SU(3)$
spontaneously broken down to $U(1)\times U(1)$. The Higgs fields
select a natural basis of simple roots. Acting with $S$-duality
on the $W$-boson states corresponding to simple roots leads to an orbit of
BPS dyon states that are magnetically charged with
respect to one of the $U(1)$'s.
The corresponding monopole solutions can be obtained by embedding $SU(2)$
monopoles into $SU(3)$ and the $S$-duality predictions reduce to the
$SU(2)$ case. Acting with $S$-duality on the
$W$-boson corresponding to a non-simple root leads to an infinite set of
new $S$-duality predictions. The simplest of these corresponds to
the existence of a harmonic form on the moduli space of $SU(3)$ monopoles
that have magnetic charge $(1,1)$ with respect to the two $U(1)$'s. We argue
that the moduli space is given by $R^3\times (R^1\times M)/\IZ$, where $M$
is Euclidean Taub-NUT space, and that the latter admits the appropriate
normalizable harmonic two form.
We briefly discuss the generalizations to other gauge groups.

\vskip .5cm
\noindent

\Date{January 1996}

\newsec{Introduction}

It is conjectured that $N=4$ supersymmetric gauge theory
is invariant under an $SL(2,\IZ)$ $S$-duality group
which includes
the interchange of strong and weak coupling
\refs{\mo\gno\bais\osborn\sen\listnew {--}\girard}.
As emphasised by Sen \sen, the conjecture makes
non-trivial predictions about the spectrum of BPS saturated dyons
which can be tested at weak coupling using semiclassical techniques.
For gauge group $SU(2)$ spontaneously broken down to $U(1)$ he showed
that the required dyon states are equivalent to the existence of
certain harmonic forms on the moduli space of classical BPS monopole
solutions. For monopole charge two he further demonstrated that the
moduli space admits the appropriate form. Evidence for the
harmonic forms for higher monopole charge has also been found \refs{\segal,
\porrati}.
The purpose of this work
is to extend these investigations to higher rank gauge groups.

For the most part we will focus on gauge group $SU(3)$ spontaneously
broken down to $U(1)\times U(1)$ but we will also discuss
how
our results apply
to other gauge groups with maximal symmetry breaking.
A basis of simple roots of the Lie algebra of $SU(3)$ is determined by the
Higgs fields \weinberg. Using this basis, the electric charge
vectors, $\vn^e$, of the $W$-bosons of positive charge
corresponding to the simple positive roots
are given by $(1,0)$ and $(0,1)$ whilst for the non-simple positive
root it is $(1,1)$. The mass and charge
of the $W$-boson states saturate a Bogomol'nyi
bound and hence they form part of a short BPS multiplet.
The magnetic duals of these states then have
magnetic charge vectors, $\vn^m$, given by the same vectors, respectively.
Weinberg has argued in \weinberg\ that the
classical monopole solutions
with $\vn^m=(1,0)$ and $(0,1)$ should be considered to be ``fundamental"
in the sense that they have a moduli space given by $R^3\times S^1$
(corresponding to translations and a dyon degree of freedom) and hence
have no ``internal" degrees of freedom. Furthermore, both the dimension
of the moduli space
of a general monopole with $\vn^m=(n_1,n_2)$,  and the BPS mass formula
are consistent with interpreting it as a multimonopole configuration
consisting of $n_1$ $(1,0)$ and $n_2$ $(0,1)$ fundamental monopoles.

Using the techniques explained in \refs{\gauntlett,\blum}
the
semiclassical quantization of the fundamental monopoles
gives rise to a BPS multiplet of states dual to the corresponding
$W$-bosons. On the other hand, the magnetic dual of the $W$-boson BPS
multiplet with charge $\vn^e=(1,1)$ must emerge as a bound state
of two fundamental monopoles. Since the monopoles with
charge $\vn^m=(1,1)$ are only neutrally stable into the decay of two
fundamental monopoles, the bound state is at threshold\footnote{$^{\S}$}{Bound
states at threshold have been recently found in the dyon spectrum
of exactly $S$-dual models with $N=2$ supersymmetry
\refs{\gauntharv,\sethietal}.}.
To determine the existence of these bound states we first
need to identify the moduli space of
monopoles with charge $\vn^m=(1,1)$.
We will argue
that is given by $R^3\times (R^1
\times M)/\IZ$ where $M$ is Euclidean Taub-NUT space. Using the results of
\refs{\gauntlett,\blum}, $S$-duality then predicts
that Taub-NUT space should admit a unique
harmonic form. We show it indeed possesses a self-dual harmonic two-form
as required.

Thus, we will show that the magnetic duals of the $W$-boson
states exist in the quantum spectrum.
Our analysis will actually go further than just checking the
electric/magnetic $\IZ_2$ subgroup
of the duality group, as we also discuss the spectrum
of dyons as well as how the
$SL(2,\IZ)$-duality predictions of the $SU(2)$ case are embedded in the
$SU(3)$ case.

The plan of the rest of the paper is as follows. In section 2 we
review some features of $S$-duality of
$N=4$ Super Yang-Mills theory, magnetic monopoles,
and the relevant parts of Weinberg's
analysis concerning the number of zero modes around a monopole
solution. Section 3 discusses in detail the $S$-duality predictions
for gauge group $SU(3)$ and contains our main results. Section
4 is a discussion section which includes some comments
concerning other gauge groups.

\newsec{N=4 Supersymmetric Gauge Theory, Monopoles  and Duality}

We consider N=4 supersymmetric Yang-Mills with arbitrary simple gauge
group. The supermultiplet includes 6 Higgs fields $\phi^I$ and
a gauge field, all taking values in the adjoint representation of the
gauge group.  The bosonic part of the action is
\eqn\bact{
S=-{1\over 16\pi}{\rm Im}\int\tau{\rm Tr}(F\wedge F+i*F \wedge F)
-{1\over 2e^2}\int \left[{\rm Tr}
D_{\mu} \phi^I D^{\mu} \phi^I+ V(\phi^I)\right]
}
where the potential is given by
\eqn\pot{
V(\phi^I) = \sum_{1\leq I < J\leq 6} {\rm Tr} [\phi^I,\phi^J]^2~,
}
and $\tau=\theta/2\pi+i4\pi/e^2$.
We have normalized the generators of the gauge group so that
${\rm Tr} t_a t_b = \delta_{ab}$.

The classical vacua of the theory correspond to solutions of the
equations
\eqn\cvac{
F_{\mu \nu} =0, \quad D_\mu \phi^I =0, \quad {\rm and} ~V(\phi^I)=0~.
}
This last equation implies that $[\phi^I, \phi^J] = 0$ for all $I,J$.
Spontaneous symmetry breaking is achieved by demanding
\eqn\ssb{
{\rm Tr} \phi^I \phi^I=v^2~,
}
as a boundary condition at infinity.
In the following we will work at a generic point
in the moduli space of vacua where the gauge symmetry is broken down
to $U(1)^l$, where $l$ is the rank of the gauge group.

A set of conserved electric and magnetic charges may be defined which
arise as central charges of the $N=4$ supersymmetry algebra:
\eqn\charges{
\eqalign{
Q^I_e &= {1\over ev}\int d{\bf S} \cdot {\rm Tr}({\bf E} \phi^I) \cr
Q^I_m &= {1\over ev}\int d{\bf S} \cdot {\rm Tr}({\bf B} \phi^I) ~,\cr}
}
where the electric field $ E_i = F_{0i}$, and the magnetic field
$B_i = 1/2 \epsilon_{ijk}F_{jk}$.
For BPS saturated states, i.e. states in the short 16 dimensional
representation of the
supersymmetry algebra, the mass is exactly given by the formula
\eqn\bpsmass{
M^2 = {v^2\over e^2}( (Q^I_e)^2 + (Q^I_m)^2\bigr)~,
}
for $\theta=0$.
A magnetic monopole solution with zero electric charge
is BPS saturated if and only if
it satisfies the Bogomol'nyi equations
\eqn\bog{
B_i =
D_i \phi~,
}
where the scalar field $\phi$ is defined by the equations
\eqn\sis{
\phi^I = \phi a^I + \hat \phi^I, \quad (a^I)^2=1~,
}
with $V(\phi^I)=0$, $D_i \hat \phi^I=0$, $a^I$ are constant and
we must apply the boundary condition \ssb\ \osborn.
For our purposes it will be sufficient to set $\hat \phi^I=0$  in the following
which allows us to focus on a single direction in the 6 dimensional Higgs
field space. BPS anti-monopoles similarly satisfy $B_i=-D_i\phi$.
Dyon states  are obtained from the monopole solutions
after semiclassical quantization.

Before proceeding further, let us review some relevant properties
of Lie algebras. The maximal abelian subalgebra will be denoted
$H$, and the $l$ generators $H_i$.
The raising and lowering operators satisfy
\eqn\raislow{
[H_i, E_\valpha] = \alpha_i E_\valpha~, \qquad [E_\valpha,
E_{-\valpha} ] = \sum \alpha^i H_i~.
}
The $H_i$ and $E_\valpha$ are linear combinations of the
generators $t_a$ previously defined.
A basis of simple roots, $\vbeta^{(a)}$ ($a=1,\cdots,l$),
may be chosen such that any
root is a linear combination of $\vbeta^{(a)}$  with  integral
coefficients all of the same sign. The term positive roots refers to
those with positive coefficients.

We may choose the Cartan subalgebra such that $\phi_0= v \vh\cdot \vH$ is
the asymptotic value of the Higgs field along the positive $z$-axis, and
$v$ is the asymptotic value of $\sqrt{{\rm Tr}\phi^2}$.
If $\valpha \cdot \vh=0$ for some root $\valpha$ then the unbroken
gauge group is nonabelian. Otherwise, maximal symmetry breaking
occurs, and $\phi_0$ picks out a unique set of simple roots
which satisfy the condition $\vh \cdot \vbeta^{(a)} > 0$ \weinberg.

The electric quantum numbers live on the $l$-dimensional root lattice
spanned by the simple roots $\vbeta^{(a)}$,
\eqn\elat{
{\bf q} = \sum n^e_a \vbeta^{(a)}~,
}
where the $n^e_a$ are integer.
The electric charge $Q^I_e$ is then given by
\eqn\elcharge{
Q^I_e=a^IQ_e~,\qquad Q_e\equiv e\vh\cdot\vq~.
}
{}For each root $\valpha$ there is a BPS $W$-boson
with $\vq=\valpha$. From \bpsmass\ we see that the $W$-bosons
corresponding to simple roots are stable, whilst those
corresponding to the non-simple roots are only neutrally stable.

Now we consider magnetic quantum numbers which arise from
topologically nontrivial field configurations. For any
finite energy solution, asymptotically we have
\eqn\bfield{
B_i = {r_i \over {4\pi r^3}} G(\Omega)~,
}
where $G$ is covariantly constant, and takes the value $G_0$
along the positive $z$-axis. The Cartan subalgebra may
be chosen so that $G_0 = \vg \cdot \vH$. This quantity
must satisfy a topological quantization condition \refs{\gno,\engwind}
\eqn\cquant{
e^{i G_0} = I~.
}
The solution to this equation is
\eqn\gis{
\vg = 4 \pi \sum n_a^m \vbeta^{(a)*}~,
}
where the $n_a^m$ are integers and the $\vbeta^{(a)*}$ are the
duals of the simple roots, defined as
\eqn\simdual{
\vbeta^{(a)*} = { \vbeta^{(a)} \over \vbeta^{(a)2} }~.
}
For maximal breaking, all the $n_a^m$
are conserved topological charges, labeling the homotopy class of the
Higgs field configuration \refs{\godol \taubes {--}\horraw}.
The magnetic quantum numbers thus live
on the lattice spanned by the $\vbeta^{(a)*}$.

The topological charge ${\bf g}$ is related to the charge $Q^I_m$
defined above by the formula
\eqn\ttfo{
Q^I_m=a^IQ_M~,\qquad Q_M\equiv {1\over e}\vg\cdot\vh ~.
}
Substituting this into the BPS mass formula \bpsmass\ we deduce
that the monopoles corresponding to the duals of the simple
roots will be stable. Those corresponding to the duals
of the nonsimple roots are neutrally stable with respect
to decay to simple root monopoles.

A general state may be labeled by the integer
valued $l$-vectors $\vn^e$ and $\vn^m$.
{}For a BPS state the mass is given by the BPS mass formula \bpsmass\ which,
using \elcharge\ and \ttfo,  can be recast
in the form
\eqn\bpsagain{
M=v|(\vh\cdot\vbeta^{(a)})n^e_a+\tau(\vh\cdot\vbeta^{(a)*})n^m_a|~,
}
where we have reinstated $\theta$.
The action
of $SL(2,\IZ)$ duality on such a state is given by
\eqn\dual{
\eqalign{(\vn^m,\vn^e)&\to(\vn^m,\vn^e)M^{-1}~,\cr
\tau&\to{a\tau+b\over c\tau +d}~,\cr}}
where $M=\bigl(\matrix{a&b\cr c&d}\bigr)$ and $ad-bc =1$, with
$a,b,c,d$ integers.
$S$-duality is generated by $S:\tau\to-1/\tau$ and $T:\tau\to\tau+1$.
When we act with $S$ we must replace the group $G$ with its dual group $G^*$
\gno. For simply laced groups the $N=4$ supersymmetric Lagrangian with
gauge group $G$ is the
same as that of $G^*$ since all fields are in
the adjoint representation\footnote{$^\dagger$}{For non-simply-laced groups
this is not true since for example $SO(2N+1)^*=SP(N)$. In this
case one does not expect the theory to be invariant under the
full $SL(2,\IZ)$ duality group, but rather a $\Gamma_0(2)$
subgroup \girard. We restrict our considerations to simply-laced
gauge groups in the following.}.
Starting with the $W$-boson states $S$-duality predicts an
infinite number of dyon BPS states.
Since the quantum moduli space is
assumed to be the same as the classical moduli space, these states should
exist for all values of the coupling constant $\tau$ and in particular
for weak coupling where semiclassical techniques are reliable.

To determine the semiclassical dyon spectrum one needs to know the
moduli space of classical BPS monopole solutions.
Using an index theorem Weinberg has argued that the moduli
space of monopoles of charge $\vn^m$ has dimension
\eqn\nzero{
d=4\sum_a n_a^m~.
}
A number of explicit monopole solutions can be constructed by embedding $SU(2)$
monopoles as follows \bais. Let $\phi^s$, $A_i^s$ be an $SU(2)$ monopole
solution with charge $k$ and Higgs expectation value $\lambda$. If we
let $\valpha$ be any root satisfying $\valpha\cdot\vh>0$ then we can
define an $SU(2)$ subgroup with generators
\eqn\sutwo{
\eqalign{
t^1 &= (2 \valpha^2 )^{-1/2} ( E_{\valpha} +
E_{-\valpha}) \cr
t^2 &= -i(2 \valpha^2 )^{-1/2} ( E_{\valpha} -
E_{-\valpha}) \cr
t^3 &= (\valpha^2 )^{-1} \valpha \cdot \vH ~.\cr}
}
A monopole with magnetic charge
\eqn\mch{
\vg=4\pi k\valpha^*
}
is then given by
\eqn\sutsol{
\eqalign{
\phi &= \sum_s \phi^s t^s + v( \vh - { {\vh\cdot \valpha }\over
{\valpha^2}} \valpha ) \cdot \vH \cr
A_i &= \sum_s A_i^s t^s \cr
\lambda&=v\vh\cdot\valpha~.\cr
}}
Since the moduli space of $SU(2)$ monopoles with charge $k$ has dimension $4k$
these solutions provide a $4k$ dimensional submanifold of monopoles
with charge \mch.
Note that by embedding an $SU(2)$ monopole with charge one we obtain
spherically
symmetric monopole solutions.

Weinberg has shown that
there is a distinguished set of $l$
``fundamental monopoles" with $\vg=4\pi\vbeta^{(a)*}$ i.e., they have magnetic
charge vectors $\vn^m$ consisting of a one in the $a$th position and
zeroes elsewhere. The reason for calling them fundamental is
twofold. Firstly, they have no ``internal" degrees of freedom:
all of these solutions can be constructed by
embedding an $SU(2)$ monopole of unit charge using
the corresponding simple
root and consequently
they have only four zero modes: three translation
zero modes and a $U(1)$ phase zero mode corresponding to dyonic excitations
of the same $U(1)$ as where the magnetic charge lies\footnote{$^*$}{One can
check that the embedded $SU(2)$ solutions are invariant under
gauge transformations of the other $U(1)$'s.}.
Secondly, a general monopole with charge $\vn^m$
can be considered to be a multimonopole configuration consisting
of $n_a^m$ monopoles of type $a$. The Bogomol'nyi mass formula \bpsagain\
and the
dimension of the moduli space \nzero\ support this interpretation.

Note that for magnetic monopoles with charge vector $\vg=4\pi k\vbeta^{(a)*}$
i.e. consisting of $k$ fundamental monopoles of the same type, the dimension
of moduli space is $4k$. Thus we deduce that these solutions can all be
obtained by embedding $SU(2)$ monopoles of charge $k$, using the embedding
based on the same simple root.

\newsec{Duality and $SU(3)$ Dyons}

In order to simplify the notation we
now restrict ourselves to gauge group $SU(3)$. The generalization of
the discussion to other gauge groups
should be reasonably clear and we will return to this in the
discussion section.

The $W$-boson states have electric charge vectors $\vn^e$=$\pm(1,0)$,
$\pm(0,1)$ and $\pm(1,1)$ corresponding to the two simple roots and
the non-simple root, respectively. Note that the $(1,1)$ $W$-boson
is only neutrally
stable into the decay of two simple root $W$-bosons.
Starting with these BPS saturated states, $SL(2,\IZ)$-duality
generates an orbit of electric and magnetic charge vectors
$(\vn^m,\vn^e)$ given by
\eqn\orb{
\bigl(p(1,0),q(1,0)\bigr), \qquad \bigl(p(0,1),q(0,1)\bigr),
\qquad \bigl(p(1,1),q(1,1)\bigr)~,
}
for relatively prime integers $p$ and $q$.
At weak coupling these states should be visible in the semiclassical
quantization of the monopole solutions. A monopole solution with charge
vector $\vn^m=(n_1,n_2)$
can be interpreted as being a multimonopole configuration consisting
of $n_1$ fundamental monopoles of type $(1,0)$ and $n_2$ fundamental
monopoles of type $(0,1)$.

First let us consider the monopoles with $\vn^m=(p,0)$ or $\vn^m=(0,p)$
i.e. $p$ fundamental monopoles of the same type.
As we mentioned at the end of the last section, all of these $SU(3)$ monopoles
can be obtained by embedding $SU(2)$
monopoles with charge $p$ using the appropriate simple root.
The dyonic states with charges $( p(1,0), q(1,0))$
and $(p(0,1),q(0,1))$ predicted by duality should thus emerge from
the semiclassical quantization of
$SU(2)$ monopoles \refs{\sen,\segal}. In this way, the predictions
of $S$-duality for gauge group $SU(2)$ are embedded in
the $SU(3)$ case.

The new predictions for $SU(3)$ monopoles thus arise in the sectors
with both magnetic quantum numbers non-zero. In particular, the
$(p(1,1), q(1,1))$ dyon states should arise as bound states of $p$
$(1,0)$ and $p$ $(0,1)$ monopoles. Note from the BPS mass
formula that these states are only neutrally stable and consequently
they should emerge as bound states at threshold. In the following, we will
prove the existence of these states for the case $p=1$.

To proceed we must quantize
the collective coordinates of the 2-monopole solution corresponding
to a $(1,0)$ and a $(0,1)$ monopole. This moduli space does
not appear to have been studied in the literature, so our
first task will be to determine its form.
By factoring out the center of mass, we expect that the moduli
space is of the form $R^3\times S^1\times M$.
The $S^1$ should correspond
to the overall $U(1)$ charge aligned along the $(1,1)$ direction.
We will see later that this is not quite correct and we must replace the
$S^1$ with $R^1$ and also make a discrete identification by $\IZ$.
{}From \nzero\ we deduce that the ``relative moduli space", $M$,
is four-dimensional.
Since the low-energy dynamics of the monopoles is given by an
$N=4$ supersymmetric quantum mechanics on
the moduli space \refs{\gauntlett,\blum}, $M$ must be hyperk\"ahler.
The spherically symmetric $(1,1)$ monopole solution obtained by
the $SU(2)$ embedding based on the non-simple positive root
corresponds to a single point in $M$.

$M$ also should admit various isometries: as there are two independent
$U(1)$ charges which are good quantum numbers $M$ should admit a
$U(1)$ group of isometries. In addition there are
isometries arising from the the action of spatial rotations,
combined in general with a gauge transformation.
One might naively think that this gives rise to an $SO(3)$ group of
isometries. However, by studying the behavior of the zero modes
about the spherically symmetric $(1,1)$ monopole solution
we will now argue
that the group of isometries is in fact $U(1)\times SU(2)$.

Let us work in $A_0^a=0$ gauge and consider the solution to the
Bogomol'nyi equation $B_i = D_i \phi$ corresponding to the embedding
of the 't Hooft-Polyakov $SU(2)$ monopole in
a nonsimple $(1,1)$ root of $SU(3)$. The zero modes
of this solution satisfy the linearized equation
\eqn\bzerom{
D_i \delta \phi + [\delta A_i,\phi]-
\epsilon_{ijk} D_j \delta A_k = 0~.
}
The gauge
condition
\eqn\bgauge{
D_i \delta A_i +  [\phi, \delta \phi] =0~,
}
is imposed to remove modes corresponding to residual gauge transformations.
As noted in \refs{\brown,\mottola}, we may combine $\delta \phi$ and
$\delta A_i$ into a 4-vector. Using the decomposition $O(4)= SU(2)\times SU(2)$
this 4-vector may be represented as a $(\half, \half)$ representation
of $SU(2)\times SU(2)$, i.e. as the $2\times 2$  matrix
\eqn\psimat{
\psi = I \delta \phi + i \sigma_j \delta A_j~,
}
where the $\sigma_j$ are the Pauli matrices. For real zero modes we
deduce that $\psi$ satisfies the reality constraint
$\psi^*=\sigma_2\psi\sigma_2$. In this notation,
equations \bzerom\ and \bgauge\ take the simple form
\eqn\psieqn{
-i \sigma_j D_j\psi + [\phi,\psi] =0~.
}
Noting that this equation does not mix the two columns of the matrix
$\psi$, we see $\psi$ may be constructed from solutions to the
2-component spinor equation
\eqn\chieqn{
-i \sigma_j D_j\chi + [\phi,\chi] =0~.
}
The spinor $\chi$ transforms as the $(\half,0)$ rep of  $SU(2)\times SU(2)$,
as does $\sigma_2 \chi^*$.
Spatial rotations correspond to the diagonal $SU(2)$ subgroup
of $SU(2)\times SU(2)$. Setting
\eqn\chiis{
\chi = \bigl( \matrix{a \cr b } \bigr)~,
}
a solution $\psi$ is
\eqn\psiis{
\psi = \chi \otimes   \bigl( \matrix{1 & 0 } \bigr) -i
 \sigma_2 \chi^* \otimes \bigl( \matrix{0 & 1 } \bigr) ~.
}
Another linearly independent solution is obtained by replacing
$\chi$ by $i\chi$.

The solutions to the Dirac equation \chieqn\ are discussed in \weinberg.
The modes are categorized by their quantum numbers with respect to
an $SU(2)$ isospin ${\bf t}$ and a $U(1)$ hypercharge $y$.
Generators of $SU(3)$ which lie in the Cartan subalgebra are
isospin singlets with $y=0$. The roots have
\eqn\isoch{
\eqalign{
t_3 E_\valpha &= { {\vbeta \cdot \valpha} \over {\vbeta^2}} E_{\valpha} \cr
y E_\valpha &= ( { {\vh\cdot \valpha} \over {\vh\cdot \vbeta} }
-t_3) E_{\valpha}~, \cr}
}
where $\vbeta$ is the root used to embed the $SU(2)$ solution.
The adjoint of $SU(3)$ decomposes as ${\bf 8}\to {\bf 3}+{\bf 2}+{\bf 2}+{\bf
1}$
with respect to $\bf t$ and the hypercharge depends on the Higgs field.
The number of normalizable modes is as follows \weinberg:
\eqn\modenum{
\eqalign{
t= \half :\quad & 0\leq |y| < \half,~ {\rm one}, \cr
& \half \leq |y|, ~{\rm zero}, \cr
t= 1 :\quad & 0\leq |y| < 1,~ {\rm two}, \cr
& 1 \leq |y|, ~{\rm zero}. \cr}
}
We will not consider the cases $t=|y|=\half$ and $t=|y|=1$ in the following
since they do not occur for maximal symmetry breaking.
Let us define the generator of spatial rotations ${\bf j} = {\bf L}+
{\bf s}$. The zero modes are eigenvectors under the combined
rotation and gauge transformation generated by ${\bf J} = {\bf j}+{\bf t}$.
Note that the $SU(2)$ embedded solution itself is spherically
symmetric with respect to this $SU(2)$.
Since the bosonic zero modes are constructed as a
tensor product of $\chi$ with a constant $(0,\half)$ spinor
(plus a piece with $-i\sigma_2 \chi^*$ tensored with another $(0,\half)$
spinor) and similarly for $i\chi$, we see that $\psi$ transforms as
a ${\bf J} \otimes {\bf \half}$
representation.

The fermion zero modes arising from the triplet of $SU(2)$
have quantum numbers $t=1$, $y=0$ and $J=1/2$. The
bosonic zero modes thus transform as a ${\bf 1}\oplus {\bf 0}$ rep
with respect to ${\bf J}$. This corresponds to the
$R^3 \times S^1$ factor of the moduli space, which arises
from the three translation zero modes of the center of mass,
together with an overall $U(1)$ phase degree of freedom.

When $\beta$ is a simple root, the two doublets with $t=1/2$ have $|y|>1/2$
and lead to no further zero modes. On the other hand
when $\beta$ is a nonsimple root the two doublets have $0\le|y|<1/2$
(with opposite signs of $y$) and there are an additional pair of
normalizable fermion zero modes.
The bosonic zero modes
then transform as two doublets with respect to $J$.
The moduli space we are interested in therefore
has a $SU(2)$ subgroup of isometries rather than the
usual $SO(3)$ group, since the zero modes are sensitive to
the center of $SU(2)$.

So we are led to look for a four dimensional hyperk\"ahler manifold with
an $SU(2)\times U(1)$ group of isometries. Moreover, there should be
a fixed point, a ``NUT", of the $SU(2)$ action corresponding
to the spherically symmetric embedded $SU(2)$ solution.
In addition, since the three complex structures are inherited from
those on $R^4$ (see e.g., \gauntlett), they should transform as a triplet under
the action of $SU(2)$.
A classification of such spaces has been carried out by Atiyah and Hitchin
\atiyah. Assuming that the manifold is complete
we are led to one of two
possibilities: Euclidean positive mass Taub-NUT space and
$R^4$. Note that the Atiyah-Hitchin
manifold is excluded both because it has just $SO(3)$ isometry and
the only fixed point set is a two-dimensional ``bolt". Likewise,
$R^3 \times S^1$ is ruled out because its isometry group
includes $SO(3)$ rather
than $SU(2)$.
Asymptotically, we expect the manifold to approach
$R^3 \times S^1$, with the $S^1$ arising from the relative $U(1)$ orientation
of the two monopoles, and the $R^3$ from
the separation of the monopole centers. $R^4$ is thus ruled out
and Taub-NUT is the unique solution.

The metric for Taub-NUT space is given by
\eqn\tnut{
ds^2=V^{-1}dr^2+V^{-1}r^2((\sigma_1^R)^2+(\sigma_2^R)^2)+V(\sigma_3^R)^2~,
}
with
\eqn\vee{
V^{-1}=1+{1\over r}~,
}
where we have scaled out the positive mass parameter and set it equal to 1/2,
and where $\sigma_i^R$ are a basis of left-invariant one-forms on $S^3$
whose explicit form is
\eqn\liof{
\eqalign{\sigma^R_1&= -\sin\psi d\theta+\cos\psi \sin\theta d\phi\cr
\sigma^R_2&= \cos\psi d\theta+\sin\psi \sin\theta d\phi\cr
\sigma^R_3&= d\psi+\cos\theta d\phi~.\cr}}
By introducing the coordinate $R=2{\sqrt r}$ it is straightforward to show
that the metric is non-singular at $r=0$ if the period of $\psi$ has period
$4 \pi$. Choosing this period, Taub-NUT has $SU(2)$ isometry. 
The generators of the $SU(2)$
isometry are given by the ``left vector fields", $\xi^L_i$, $i=1,2,3$, whose
explicit form can be found, for example, in
\gauntharv. The generator of the extra $U(1)$
isometry is the ``right vector field" $\xi^R_3=\p/\p_\psi$.

To ensure that
semiclassical quantization yields a spectrum of dyons with
electric charges lying on the lattice \elat, we must replace the overall
$S^1$ factor by $R^1$ and perform a discrete identification by $\IZ$, as we now
explain \lwy. 
The electric charge operators $Q_1$ and $Q_2$ defined by
\eqn\chop{
Q_1 = {2\over 3} (2\vbeta^{(1)} + \vbeta^{(2)} )\cdot \vH ~,\qquad
Q_2 = {2\over 3} (\vbeta^{(1)} +
2 \vbeta^{(2)} )\cdot \vH ~,
}
take integer values $n_a^e$. 
The total electric charge $Q_\chi$,
conjugate to the collective coordinate $\chi$, can be determined in terms
of the $Q_i$ by expanding the Bogomol'nyi mass formula \bpsagain\ for 
$\vn^m=(1,1)$ and (small) general $\vn^e$. 
The kinetic energy terms in the Hamiltonian for the
collective coordinates conjugate to the charges $Q_i$ can
also be determined
from \bpsagain\ by considering two well separated fundamental monopoles.
The relative 
electric charge $Q_\psi$ conjugate to $\psi$ is then determined
by reexpressing these kinetic energy terms in terms of $Q_\chi$ and
$Q_\psi$ using the fact that the moduli space is in a factored form.
We find
\eqn\torop{
Q_\chi = (m_1 Q_1+m_2 Q_2)/(m_1+m_2) ~, \qquad
Q_\psi = \half(Q_1-Q_2)~,
}
where $m_a= v (4\pi/g^2)\vbeta^{(a)} \cdot \vh$  is the mass of
the $a$th (pure) fundamental monopole.
It now follows that we must
impose the discrete identification
$(\chi,\psi) \sim (\chi+2\pi, \psi+4\pi m_2/(m_1+m_2))$, 
to reproduce the allowed lattice of charges \elat. In general, the
coordinate $\chi$ is not periodic and hence the moduli space
is given by $R^3 \times
(R^1 \times M)/\IZ$.
Note that if $m_2/m_1=p/q$ is rational then
$\chi$ has period $2\pi(p+q)$ and the moduli space is then $R^3 \times
(S^1\times M)/\IZ_{p+q}$.

The low-energy dynamics of the monopoles is given by an $N=4$
supersymmetric quantum mechanics on the moduli space. As discussed in
\refs{\gauntlett,\blum} the states
are given by differential forms on moduli space. The Hamiltonian is the
Laplacian acting on these forms. The basis of
16 forms on $R^3\times R^1$ leads to a BPS supermultiplet of 16 states
with total charge $Q_\chi=x$, $x\in R^1$. These should be tensored with forms
on Taub-NUT space with relative charge $Q_\psi=n/2$, $n\in \IZ$, 
where $x+ n m_2 /(m_1+m_2)\in \IZ$ 
to ensure the state is well defined on the moduli space. The value of $n$
is simply the eigenvalue of the operator $-i\xi_3^R$.

Recall that $S$-duality predicts that there should be a tower of
dyon BPS states with magnetic charge $(1,1)$ and electric charge $q(1,1)$,
for arbitrary integer $q$.
In order for these states to exist there must be a
normalizable harmonic form
on $M$.
It should be harmonic to ensure that when it is combined with
the forms on $R^3\times R^1$, the state
saturates the Bogomol'nyi bound as explained in more detail in \gauntlett.
Since the electric charge of the predicted states is only in
the $(1,1)$ direction,
the harmonic form should carry no
relative electric charge which means that it must be invariant under
the generator of the $U(1)$ isometry $\xi^R_3$. The above
discussion then implies that
$x$ is integer valued and that 
$\vn^e=x(1,1)$ as required.
Since we only require a single harmonic form,
it should either be self-dual or anti-self dual.
Taub-NUT space does indeed admit a harmonic self-dual form
given by\footnote{$^\ddagger$}{The existence of this form was noted
in a different context in \refs{\brill,\pope}.}
\eqn\form{
\omega={r\over r+1}\sigma^1\wedge\sigma^2+{1\over (r+1)^2}dr\wedge
\sigma_3=d(V\sigma_3)~.}
It is straightforward to check that it is well defined at $r=0$,
is normalizable and
is invariant under the action of
$\xi^R_3$, exactly as required by duality.

\newsec{Discussion}

Starting with the charged $W$-boson BPS states we have argued that
$S$-duality for gauge group $SU(3)$
predicts an infinite tower of BPS states with electric and
magnetic charge
vectors given by \orb. Making the weak assumption that the
$W$-boson states are the only
purely electric charged states in the theory, $S$-duality implies
that these are the only BPS states in the theory
whose electric and magnetic charge
vectors are parallel.
If other states of this type existed, then
acting with $S$ duality transformations would give a purely electric charged
state which was not a $W$-boson. Note that if the charge vectors are not
parallel then there may exist states which break one quarter of the
supersymmetry
and form medium size supermultiplets. Such solutions have been found
in supergravity (see for example \cvetic) and it would be interesting to know
if they also existed in field theory.

We have shown that the existence of the
states in \orb\ with magnetic charge vector $(p,0)$ or
$(0,p)$ is equivalent to their existence in the $SU(2)$ theory.
The main result of the paper was the verification that the
states with magnetic charge vector $(1,1)$ also exist.
The general states with charges $(p(1,1),q(1,1))$ should emerge via
the existence of certain harmonic forms on the moduli space of
$p$ $(1,1)$ monopoles. An explicit verification of this
seems a difficult undertaking at present since the moduli spaces
are not known.

The preceding discussion  for $SU(3)$
has a simple extension to a more general simply-laced gauge
group. The W-bosons, with electric quantum numbers $\vn^e$,
generate an orbit of electric and magnetic quantum numbers
given by
\eqn\chorb{
(\vn^m,\vn^e)=( p \vn^e, q \vn^e)~,
}
where the integers $p$ and $q$ are relatively prime.
Exact duality for gauge group $SU(2)$ implies the existence of bound states
when $\vn^e$ is that of a simple root W-boson.
As before, the new predictions arise
when one considers a nonsimple root $\valpha$.
The analysis of section 3 for $\vn^m=(1,1)$ carries over directly
when $\valpha^* = \vbeta_1^* + \vbeta_2^*$, with
$\vbeta_1$ and $\vbeta_2$ different simple roots, and
when the regular embedding of the $SU(2)$ monopole solution
using the root $\valpha$ gives rise to fields transforming
as a complex doublet of the $SU(2)$, in addition to the usual
triplet. A list of such embeddings may be extracted from
table 58 of \slansky. The moduli space of these monopole solutions
will be the same and the required bound state will exist, as discussed above.

More generally, the $SU(3)$ predictions for $\vn^m=p(1,1)$
will be embedded in higher
rank gauge groups. However, since some of the nonsimple roots must
be expressed as a sum over $n$ ($n>2$) simple roots, there will also be new
predictions. $SL(2,\IZ)$ invariance predicts that the corresponding
moduli spaces also admit unique harmonic forms. As far as we know, these
moduli spaces are also not yet known.

\bigskip
{\bf Acknowledgements}

We thank  M. Cederwall, S. Chaudhuri, F. Dowker, G. Gibbons,
J. Harvey, J. Schwarz,
A. Sen, A. Strominger, L. Thorlacius, E. Weinberg
and especially E. Witten for useful discussions.
We also thank the Aspen
Center for Physics for hospitality during
the initial stages
of this project. This work was supported in part by
NSF Grant PHY 91-16964 and by the U.S. Dept. of Energy
under Grant No. DE-FG03-92-ER40701.

\bigskip
{\bf Note Added:} After this work was finished we learnt of related work
of by Lee, Weinberg and Yi \lwy\ and by Connell \con.

\listrefs
\end